\documentstyle[psfig,12pt]{article}

\newcommand{\ptitle}{The El Ni\~{n}o Stochastic Oscillator}
\newcommand{\pauthors}{Gerrit Burgers}
\newcommand{\paffil}{
Royal Netherlands Meteorological Institute, The Netherlands}  

\begin{document}
\pagestyle{empty}
\bibliographystyle{agunice}
\begin{titlepage}
\mbox{}\\[15mm]
\begin{center}
\LARGE\bf \ptitle\\[20mm]
\large \pauthors\\
\normalsize\it \paffil\\[10mm]
\large
Submitted to Journal of Climate\\
         April, 1997\\[30mm]
\end{center}

\begin{center}
\parbox{0.80\textwidth}{
\begin{tabular}[t]{l}
Gerrit Burgers\\
Oceanographic Research Division\\
Royal Netherlands Meteorological Institute (KNMI)\\
PO Box 201, 3730 AE  De Bilt\\
The Netherlands\\
\end{tabular}
\begin{tabular}[t]{ll}
Phone: & +31 302206682\\
fax:   & +31 302210407\\
e-mail:& burgers@knmi.nl
\end{tabular}
}
\end{center}
\end{titlepage}

\pagebreak
\vspace*{20cm}

\pagebreak

\setcounter{page}{1}
\mbox{}\\[5mm]
\begin{center}
\Large\bf \ptitle\\[3mm]
\normalsize\sc \pauthors\\[3mm]
\small\it \paffil\\[7mm]
\normalsize \bf Submitted to Journal of Climate, April 1997   \\[7mm]
\end{center}

\pagestyle{plain}

\begin{abstract}

Anomalies during an El Ni\~{n}o are dominated
by a single, irregularly oscillating, mode.  
Equatorial dynamics has been linked to delayed-oscillator models
of this mode.
Usually, the El Ni\~{n}o mode is  regarded as an unstable mode of the 
coupled atmosphere system and the irregularity is attributed to noise 
and possibly chaos.

Here a variation on the delayed oscillator is explored.
In this stochastic-oscillator view,
 El Ni\~{n}o is a stable mode excited by noise.  

It is shown that the autocorrelation function of the observed NINO3.4 
index is that of a stochastic oscillator, within the measurement 
uncertainty.

Decadal variations as would occur in a stochastic oscillator are shown 
to be comparable to those observed, only the increase in the 
long-term mean around 1980 is rather large.

The observed dependence of the seasonal cycle on the variance    and 
the correlation is so large that it can not be attributed to the
natural variability of a stationary stochastic oscillator.
So the El Ni\~{n}o stochastic-oscillator parameters must depend
on the season.

A forecast model based on the stochastic oscillator
with a variance  that depends on the season has a skill 
that approaches that of more comprehensive statistical models:
over the period 1982-1993, 
the anomaly correlation is 0.65 for two-season lead forecasts.

\end{abstract}

\section{Introduction \label{sec:INT}}

ENSO indices are highly correlated.  
The correlation between the 
Southern Oscillation and El Ni\~{n}o has given even ENSO its name.

Another well-known example is the correlation between sea surface  
height (SSH) and sea surface temperature (SST) anomalies in the
Eastern Pacific.  The correlation of the NCEP analyzed SSH anomaly 
(Ming Ji, private communication) and the observed SST anomaly in the 
NINO3 region exceeds 0.85 for the period 1980--1995.  The same applies 
for the correlation between the observed SSH anomaly at Santa Cruz 
(as made available by the University of Hawaii Sea Level Center) 
and observed NINO3 SST anomalies over the period 1978--1995. 
In general, it seems that 
ENSO related anomalies are dominated by one single mode.

It is widely accepted that
the delayed-oscillator 
mechanism of Suarez and Schopf (1987) and Battisti and Hirst (1989)  
explains this mode, or at least
the predictable part of the interannual oscillations
 (Kleeman 1993).
The main idea of the delayed oscillator is 
as follows (Cane 1992).
A positive SST anomaly in East Pacific gives rise to an eastward wind anomaly
around the equator in a region centered at about $170^\circ$W.
  This wind anomaly not only
generates an eastward propagating Kelvin wave 
that deepens the thermocline and increases the SST anomaly in the 
Eastern Pacific,
but also generates westward propagating Rossby waves.  These
reflect at the Western boundary,
and come back as a delayed Kelvin wave that decreases the 
SST anomaly in the Eastern Pacific.  Eventually the combination of 
the positive direct and 
negative delayed feedback leads to an oscillation with a period
of $\cal{O}$(3 -- 4)year, which is an order of magnitude larger 
than the delay time. 

Given this picture, the following questions should be considered: how 
does a perturbation grow initially, why is it bounded and why are 
the oscillations irregular.

Usually, see {\em e.g.} Suarez and Schopf (1987) and 
Battisti and Hirst (1989), it
is assumed that the mode simply grows because it is unstable, and that 
it is bounded because of non-linear effects.  The irregularity could 
arise from non-linearity in the ENSO dynamics as in the original 
Zebiak-Cane (1987) model, from interaction between the ENSO
dynamics and the annual cycle (Jin et al.\ 1994, Tziperman et al.\ 1994), 
or from 
``noise'' from atmospheric variability not directly related to ENSO\@.
In the latter case, Battisti and Hirst (1989) 
and Philander (1990) argue that part of the year 
the basic state could be stable (Hirst 1986) and the development of anomalies 
susceptible to external perturbations.  
Mantua and Battisti (1995) list intraseasonal 
oscillations
and the south Asian monsoon as plausible sources for atmospheric
noise in the western Pacific.
Kleeman and Moore (1997) analyze how atmospheric noise limits the
predictability of an intermediate El Ni\~{n}o model.

\begin{figure}[t]
\par
\vspace*{1mm}
\centerline{\hspace*{0mm}\hbox{
\psfig{figure=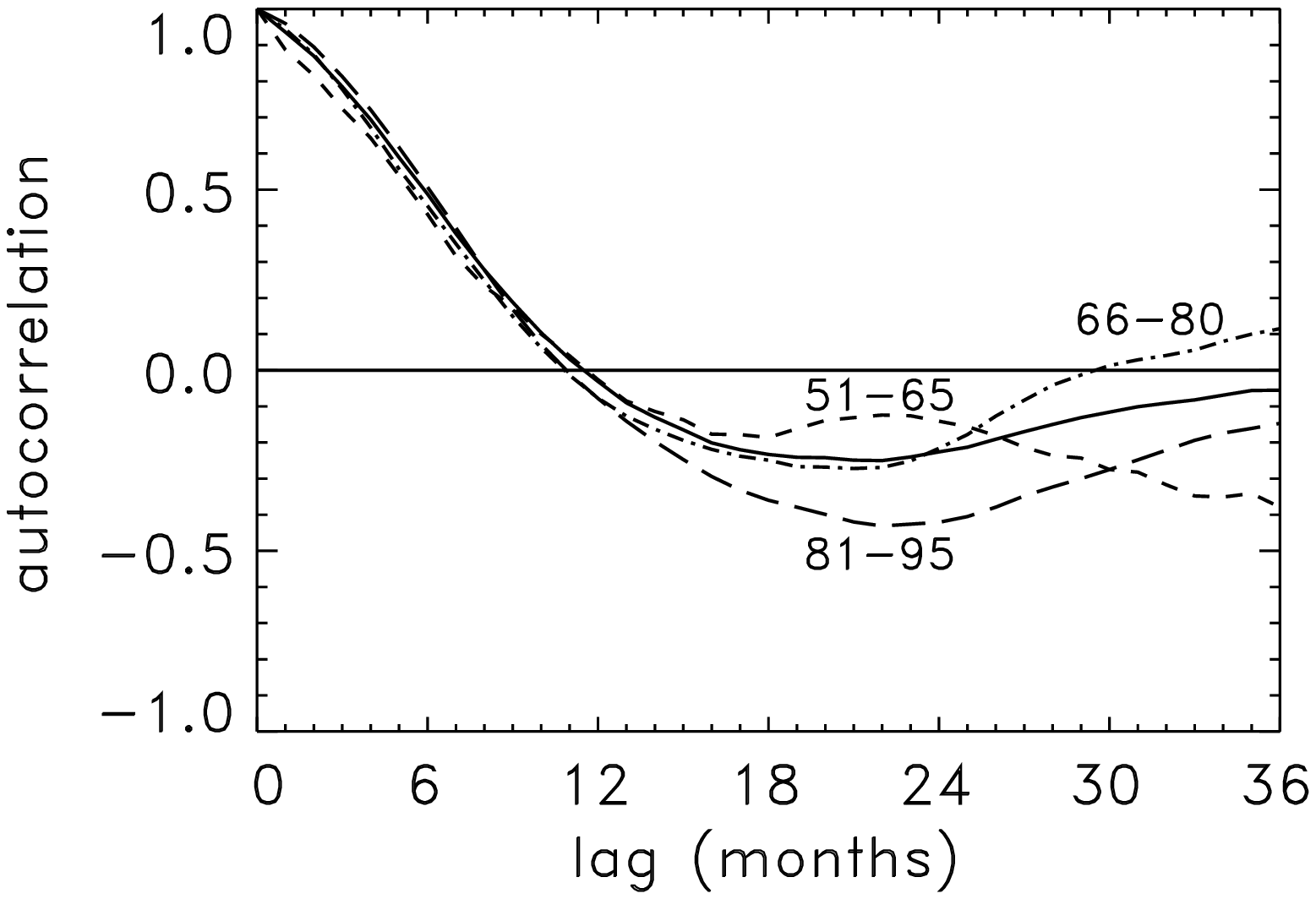,width=10.0cm,angle=0}
}}
\par
\addtolength{\baselineskip}{-2mm}
{\small
\setlength{\parindent=5mm}
\indent {\sc Fig.\ 1.}
The autocorrelation of the observed NINO3.4 index over the
period 1951--1995 (solid line) and three subperiods 
(dashed and dash-dotted lines).
}
\end{figure}


Consider now the autocorrelation function of the NINO3 or the NINO3.4
index\footnote{
The NINO3 index is the mean SST anomaly of the region
$5^\circ$S--$5^\circ$N, 
$150^\circ$W--$90^\circ$W,
and the NINO3.4 index of the region
$5^\circ$S--$5^\circ$N, 
$170^\circ$W--$120^\circ$W. }. 
The autocorrelation function of the NINO3.4 index over the period 1951--1995
is shown in Fig.\ 1. 
It is a fairly smooth function of the delay time 
and looks like a decaying cosine function.  This suggests
an alternative mechanism to explain the size and 
irregularity of ENSO\@.

In this explanation, the ENSO mode is stable, and noise 
both excites this mode and makes it irregular.
    
The simplest model which exhibits such behaviour, that of a 
stochastic oscillator, has an autocorrelation function that is a  
decaying cosine, just like observed.  
So I propose a stochastic-oscillator mechanism to describe ENSO\@.
The idea is not new.  
Jin (1997) discusses in some detail a stochastic 
oscillator as one possible mechanism to excite his coupled recharge 
oscillator ENSO model.  
New in the present paper is that the stochastic oscillator is tested
as a forecast model, and that the consequences of the
stochastic oscillator mechanism for decadal variability are
investigated.

Hasselmann (1976) introduced the concept of stochastic forcing 
as a source of variability in 
climate modelling.  Most applications have dealt with the case of a 
red variability spectrum caused by a white noise forcing.
Lau (1985) proposed that ENSO fluctuations are the result of stochastic 
forcing inducing transitions between multiple equilibrium states.
The idea of a stochastic oscillator is a very natural one too.
It has been applied to model variability of the
thermohaline circulation (Griffies and Tziperman 1995)
and to discuss decadal variability 
(Griffies and Bryan 1997; M\"{u}nnich et al.\ 1997).
A decadal delayed oscillator model for exchanges between the tropics
and the extratropics has been proposed by Gu and Philander (1997).
Chang, Ji and Li (1997) discuss stochastically excited oscillatory
modes in the context of decadal variability in the Tropical Atlantic
and Jin (1997) for ENSO\@.

Above, the delayed-oscillator picture and the observed NINO3.4 index
autocorrelation
suggested to look at a stochastic oscillator.  Another motivation 
stems from the success of hybrid coupled models 
(Balmaseda et al.\ 1994, Latif 1987, Barnett et al.\ 1993), which
have stable coupled modes.
Fl\"{u}gel and Chang (1996) and Eckert and Latif (1997) 
studied the impact of an extra stochasting forcing in a HCM.
The stochastic oscillator can be viewed as a representation of the
main coupled mode of such a HCM with noise.
Alternatively, one can arrive at the stochastic oscillator
if one uses statistical techniques for signal processing, like fitting 
autoregressive-moving average (ARMA) models to timeseries, 
or making a POP or CCA analysis of  observed fields.
Indeed, Penland and Sardeshmukh (1995), who made an inverse modeling
analysis of SST anomalies, have put forward the notion that 
the El Ni\~{n}o system is stable and driven by white noise, 
although they do not propose a system of the 
simple stochastic-oscillator type.

In section {\ref{sec:MOD}} the stochastic oscillator is formulated, 
and its autocorrelation function is compared with the observed 
autocorrelation of the NINO3.4 index.  

Simply by chance, considering a series of periods of say fifteen years,  
there will be 
fluctuations in properties like mean, variance and apperent 
autocorrelation parameters that might account for the observed 
fluctuations in these parameters.  
This will be discussed in section ~{\ref{sec:DEC}}.

In section {\ref{sec:SEA}} the influence of the seasonal cycle
is discussed.  A simple modification of the stationary 
stochastic oscillator which allows for a seasonal dependence
of the variance    is introduced.

In section {\ref{sec:FOR}} forecasts made with the stochastic 
oscillator are evaluated.  The  simple model, based on the timeseries 
of a single variable,
has a forecast skill that is comparable to that of some models 
actually used for ENSO predictions.

Finally, section {\ref{sec:DIS}} contains a discussion and the conclusion.
Some technical aspects of the stochastical oscillator 
are discussed in {\ref{sec:APP}}.

\section{The stochastic oscillator model  \label{sec:MOD}}

  The observed autocorrelation of the NINO3.4 index over the period
1951-1995 was shown in Fig.\ 1, 
toghether with the autocorrelations over the three 15-year long subperiods.
The NINO indices have been made available by the 
U.S. National Centers for Environmental
Prediction (NCEP), and are the
result of the processing of satellite, ship and buoy observations 
(Reynolds and Smith 1994). 
 
The autocorrelation has a rather regular shape, like a damped 
oscillation.  The differences 
between the three fifteen year periods shown are considerable, though.  
This suggests that random fluctuations are important,
and that it is probably not meaningful to extract more than
two or three parameters above the noise from a fit to this autocorrelation.

    As mentioned in the Introduction, the delayed-oscillator model    
is about the simplest system ENSO can be reduced to.  At the level of
simplification of the delayed-oscillator, one can just as well take
a regular oscillator as a model of the dynamics of the coupled system.
The interpretation of the delayed-oscillator is of course more direct.  
To fit the observed autocorrelation, some source of irregularity has to
be introduced.  One of the simplest ways to accomplish this is just
adding noise.  The resulting system is the stochastic oscillator.

The basic equations of the discrete stochastic oscillator can be written 
as follows:
\begin{eqnarray}
\label{eq:stoch}
     x_{i+1} & = & a x_i  - b y_i + \xi_i    \nonumber \\
     y_{i+1} & = & b x_i  + a y_i + \eta_i    \,
\end{eqnarray}
Here  $x$ and $y$ are the variables of the two degrees of freedom of the
oscillator, $a$ and $b$ are constants and $\xi$ and $\eta$ are noise
terms, which may be correlated.  It is assumed that noise between different 
timesteps is not correlated.
The constants $a$ and $b$ can be related to a 
oscillation period $T = 2 \pi  \omega^{-1}$ and a decay time scale
$D = \gamma^{-1}$ as follows:
\begin{equation}
      a + i b = e ^{- \gamma\Delta t} \, e ^{i \omega \Delta t } \; ,
\end{equation}
where $\Delta t$ is the timestep.  

Throughout this paper, the value of the timestep is $\Delta t$= 1 month.

The deterministic part of the stochastic oscillator should be stable,
{\em i.e. } $a^2 + b^2 < 1$, otherwise there would be
no bound to the amplitude of the oscillations.  If the oscillator 
is stable, then the variance of $x$ is proportional to the noise 
variance, see
   ({\ref{eq:varia}}).       

The autocorrelation function of $x$ that corresponds to 
   ({\ref{eq:stoch}}) is
\begin{equation}
\label{eq:autoco}
 \rho_k = 
                           e^{- k \gamma \Delta t} 
             \cos ( k \omega \Delta t + \alpha ) / \cos \alpha \, ,
\end{equation}
with the phase shift $\alpha$ a function of both the timescales
 $T = 2 \pi \omega ^{-1}$
and $D = \gamma^{-1}$
and the noise variances, see {\ref{sec:APP}}.

Next the identification of $x$ with an El Ni\~{n}o SST index is made.
The variable $y$ stands for another, independent index. 
One could think of $y$ as a diagnosed  El Ni\~{n}o tendency.
It will remain unidentified here.
Jin (1997) identifies equatorial zonal-mean heat-content as the second 
important ENSO variable.  However, his stochastic oscillator equations, 
which are very  much the same as (\ref{eq:stoch}), are formulated in 
terms of the East Pacific temperature anomaly and the West Pacific 
thermocline anomaly. 


\begin{figure}
\par
\vspace*{1mm}
\centerline{\hspace*{0mm}\hbox{
\psfig{figure=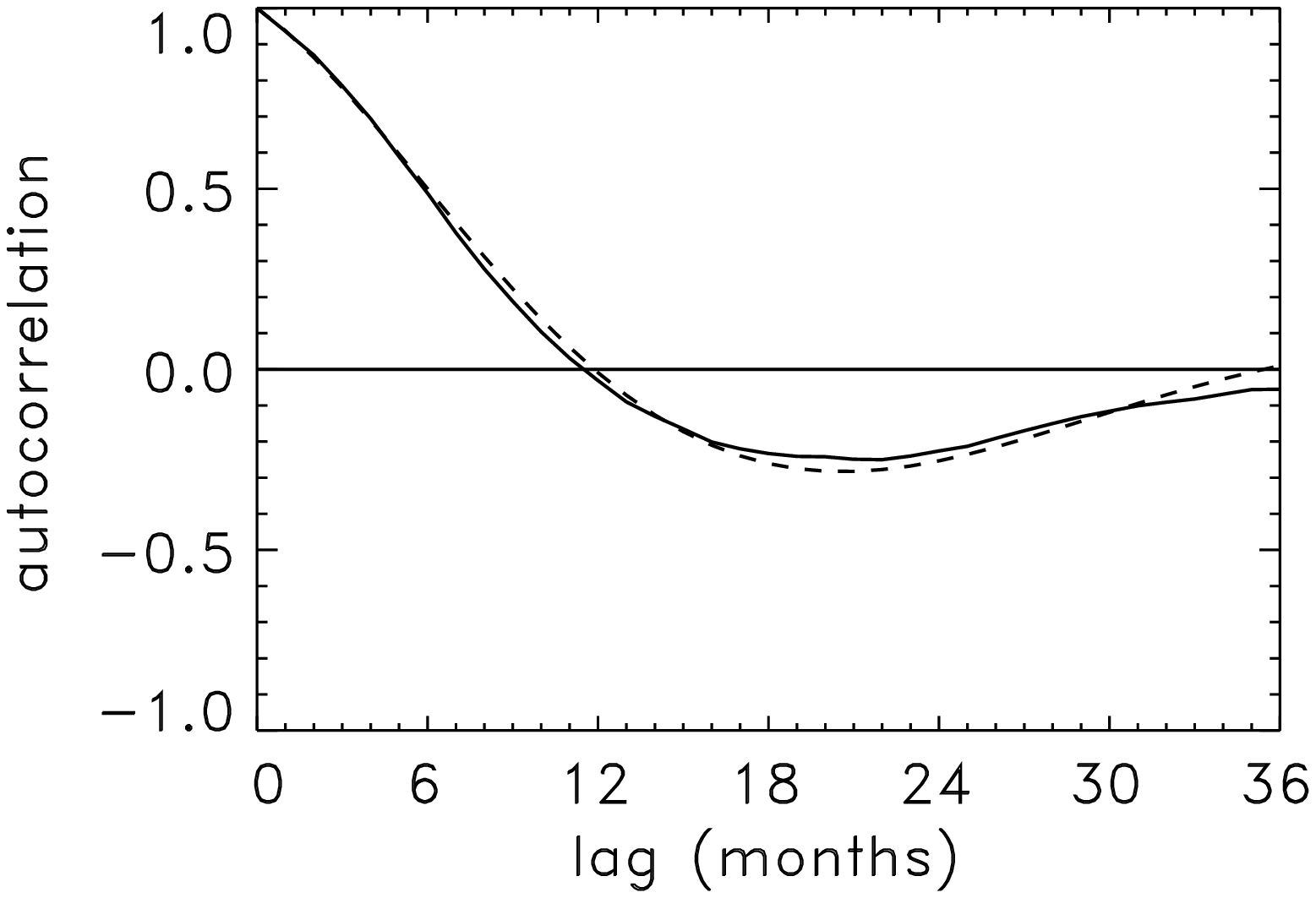,width=10.0cm,angle=0}
}}
\par
\addtolength{\baselineskip}{-2mm}
{\small
\setlength{\parindent=5mm}             
\indent {\sc Fig.\ 2.}
The autocorrelation of the observed NINO3.4 index over the
period 1951--1995 (solid line) compared to that of a
stochastic-oscillator fit (dashed line).
}
\end{figure}


If one considers $x$ only,    ({\ref{eq:stoch}}) is equivalent 
to an  ARMA(2,1) process, as shown in {\ref{sec:APP}},
\begin{equation}
\label{eq:stochx}
  x_{i+1} = 
      2 a x_i - (a^2+b^2) x_{i-1} + \epsilon_i - k \epsilon_{i-1} \; . 
\end{equation}
Note that {\em two} noise parameters enter 
({\ref{eq:stochx}}): the variance of $\epsilon$ and $k$ ($-1 \leq k \leq 1 $).

\begin{figure}
\par
\vspace*{-4mm}
\centerline{\hspace*{0mm}\hbox{
\psfig{figure=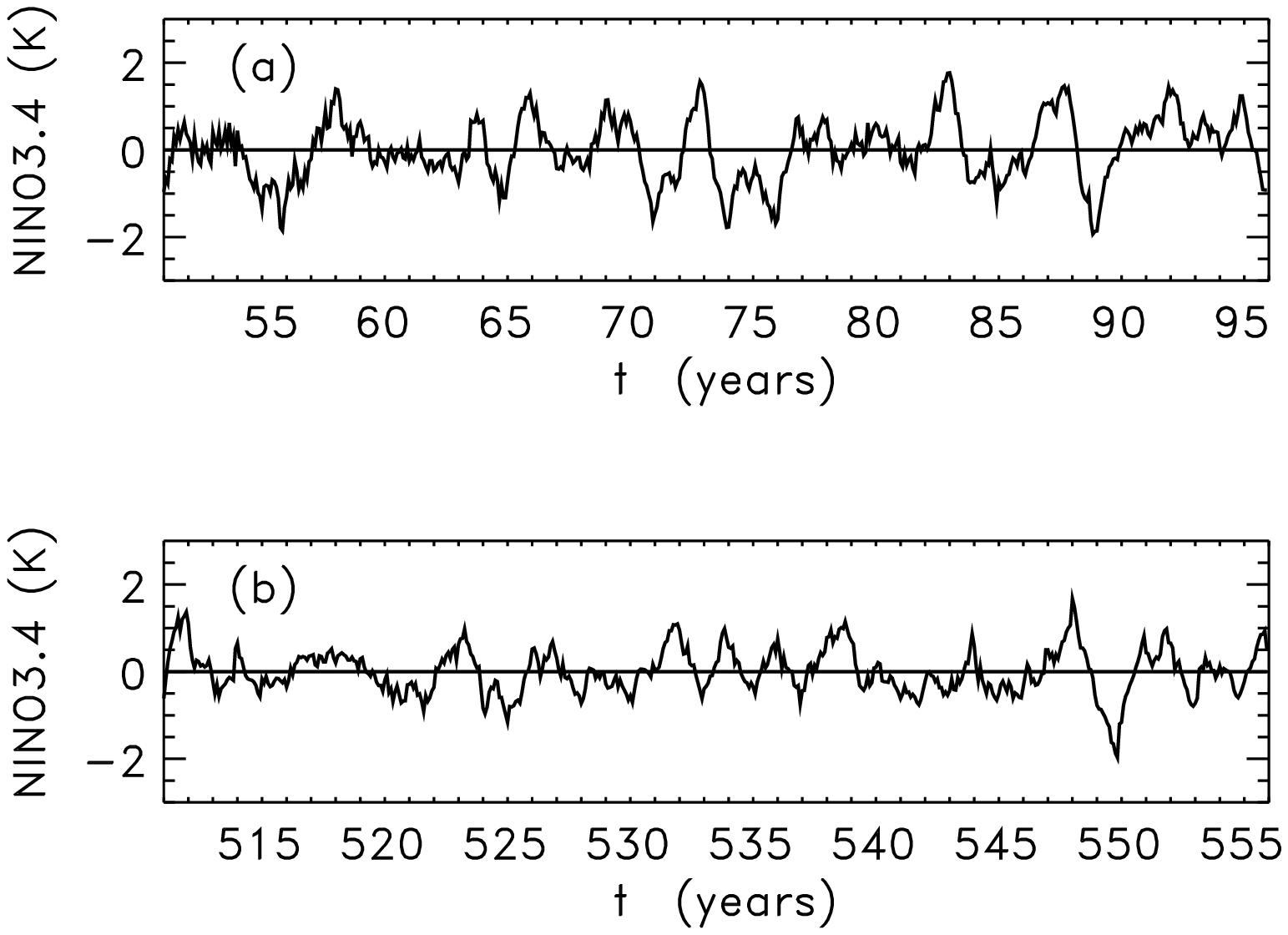,width=10.0cm,angle=0}
}}
\par
\addtolength{\baselineskip}{-3mm}
{\small
\setlength{\parindent=5mm}             
\indent {\sc Fig.\ 3.}
(a) Observed NINO3.4 index over the period 1951--1995 with respect
to the climatology over the same period.
(b) Part of a time series generated from a Monte Carlo run with a
stochastic oscillator.  
The period, decay time scale and noise parameters of the
stochastic oscillator were determined from a fit to the observed data.
}
\end{figure}


In Fig.\ 2 the autocorrelation of the observed NINO3.4 index for the
period
1951--1995 is compared to that of a stochastic-oscillator fit. 
The parameters and error estimates were obtained from a maximum-likelihood 
procedure.  Their values are
$T = 47 \pm 6$ months,
$D = 18 \pm 6$ months, and
$  k = 0.86 \pm 0.05$.  
The values of $T$, $D$, and $k$ correspond to a value of 
$\alpha = -2^\circ  \pm 7^\circ $ in
({\ref{eq:autoco}).  
The errors are highly correlated.
The variance of the driving noise is 
$\langle \epsilon \epsilon \rangle = (0.24{\rm K})^2$, 
to be compared to the rms magnitude of the signal, which is $0.7{\rm K}$.
The correspondence is good, in view of the differences between 
the observed curves in Fig.\ 1.   A better assesment of the correspondence
will be made in section {\ref{sec:DEC}}.

In Fig.\ 3, a typical Monte Carlo time series generated with the same mean 
parameters and noise is compared to a part of the observed 
NINO3.4 timeseries.

In Fig.\ 4, the residues,
{\em i.e.\ }the estimates for 
$\epsilon_i$ in    ({\ref{eq:stochx}}), are shown for the fit to
the observed timeseries over the period 1950--1995.
The r.m.s. of the autocorrelation of the residues 
for the first 24 months is $0.05$ and the largest
value is $0.10$, values that are compatible
with a white noise process for a timeseries of this length.

A similar fit can be made to the observed NINO3 time series for the
period 1951--1995.
A good fit is found for
$T  = 46 \pm 6 $ months,    
$D  = 17 \pm 5 $ months, and
$ k = 0.90 \pm 0.04$ ($\alpha = 10^\circ \pm 5^\circ$).
The variance of the driving noise is $(0.34{\rm K})^2$, 
the variance of the signal $({\rm 0.85K})^2$.  
The difference in time scales between the NINO3 and NINO3.4 cases
is  insignificant,
just as one expects if the NINO3 and NINO3.4 indices are 
manifestations of the same mode but with different noise.

     How do the above time scales fit in the delayed-oscillator equation
\begin{equation}
\label{eq:delay}  
     dT(t) / dt  = c T(t) - b T(t-\tau)
\end{equation}
of Suarez and Schopf (1988) and Battisti and Hirst (1989)?  
The delayed oscillator    ({\ref{eq:delay}}) has solutions of the form
$\exp ( -\gamma t + i \omega t )$.  For the main mode, one has
\begin{eqnarray}  
   (\gamma + c)^2  +  \omega^2  = b^2 e^{2 \gamma \tau} & & \\ \nonumber
     (\gamma + c)    =  \omega / \tan ( \omega \tau ) & &
\end{eqnarray}  
An essential element
of the delayed oscillator is that the parameters $c$ and $b$ are 
positive.
However, although a positive $c$ means a local instability, 
this does not exclude a stable oscillator
(Suarez and Schopf 1988, Battisti and Hirst 1989).  

Actually, slightly  different choices of the parameters $b$, $c$ and $\tau$ 
in the delayed oscillator ({\ref{eq:delay}}) can make
the difference between stable and unstable.
The choice $c=2{\rm yr}^{-1}$, 
$b=4{\rm yr}^{-1}$, and a delay time of $\tau=0.5{\rm yr}$,
as in Battisti and Hirst (1989), leads to
a harmonic oscillator with a period of 3 years and a {\em growth} scale
of $-\gamma=1.1{\rm yr}$, which excludes a stochastic oscillator.
Changing this to
$c=2.4{\rm yr}^{-1}$, $b=2.8{\rm yr}^{-1}$, and a delay time of
$\tau=0.3{\rm yr}$,   leads to a period of 4 years and a
{\em decay} scale of $\gamma=1.5{\rm yr}$, which is well within 
the ranges for the stochastic oscillator found above.

It is still an open question 
whether in nature the delayed oscillator is stable or unstable,
see Schneider et al.\ (1995).

\begin{figure}
\par
\vspace*{-1mm}
\centerline{\hspace*{0mm}\hbox{
\psfig{figure=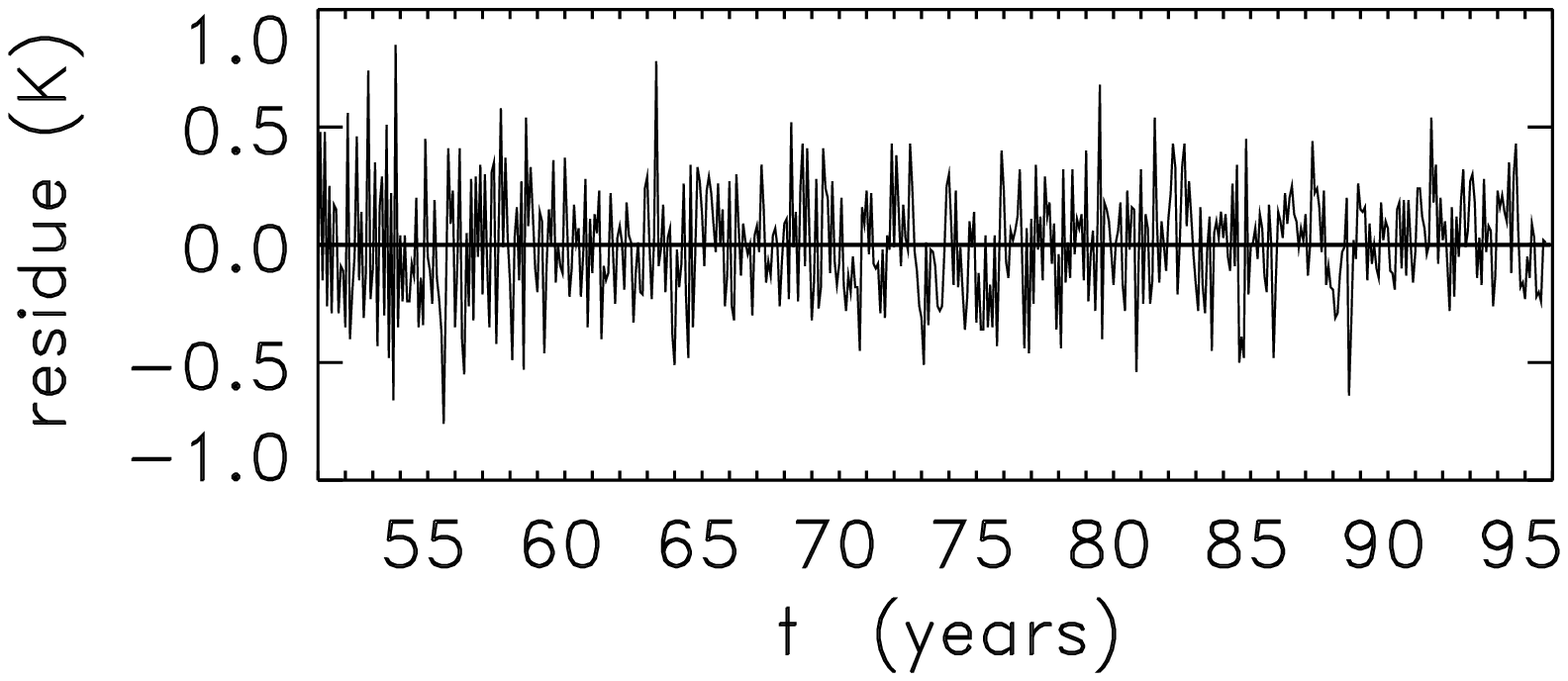,width=10.0cm,angle=0}
}}
\par
\addtolength{\baselineskip}{-2mm}
{\small
\setlength{\parindent=5mm}             
\indent {\sc Fig.\ 4.}
Residues of the stochastic-oscillator fit to the observed NINO3.4
timeseries over the period 1951-1995.
}
\end{figure}


\section{Decadal variability \label{sec:DEC}}

Over the years, properties of El Ni\~{n}o appear to have changed
from one decade to another.  {\em E.g.}, the annual NINO3.4 mean over
the fifteen year period 1981--1995 is about $0.3{\rm K}$ 
higher than the annual mean
over the period 1951--1980.  The variance and the 
autocorrelation function (see Fig.\ 1 ) differ as well.
In addition, there are appreciable differences in forecast skill.
In Table 1, these differences are summarized.  The error estimates for the
stochastic-oscillator parameters 
correspond to a decrease of 0.5 in log-likelihood with respect to
the maximum-likelihood estimate.

\begin{table}[b]
\begin{center}
\begin{tabular}{|l|lllcccc|}
\hline  
        &  \hspace*{2mm} T   &  \hspace*{2mm} D   &  \hspace*{4mm} k     & 
                   b    &$\sigma$& FS  & FP\\
\hline  

1951--65 &  $80 ^{+70}_{-20} $ &  $19 ^{+9}_{-5} $ &  $1.00^{+ 0.00}_{-0.04}$  &  0.02  &  0.60  &  5  & 4\\
1966--80 &  $39 ^{+8}_{-6}   $ &  $14 ^{+10}_{-6}$ &  
          $0.76 ^{+0.09}_{-0.12} $ &  0.03  &  0.70  &  6  & 5\\
1981--95 &  $39 ^{+6}_{-5}   $ &  $17 ^{+16}_{-8}$ & 
          $0.73 ^{+0.11}_{-0.13} $ &  0.30  &  0.77  &  7  & 5\\
1951--95 &  $47 { \scriptstyle \pm  6  }$ &  $18  {\scriptstyle \pm 6} $ &  
          $0.86 { \scriptstyle \pm 0.05}$ &  0.12  &  0.71  &  6  & 5\\
\hline  
\end{tabular}
\end{center}

\par
\addtolength{\baselineskip}{-2mm}
{\small
\setlength{\parindent=5mm}
\indent{\sc Table 1.}
Observed decadal variability of the El Ni\~{n}o NINO3.4 index.  
Period T (months), decay time scale
D (months), ARMA parameter k, bias b (K), standard deviation $\sigma$ (K),
and lead FS (months) at which the anomaly correlation skill of 
stochastic-oscillator forecasts drops to $0.6$, and the same for
persistence forecasts (FP).
}

\end{table}

Here the question will be addressed how much of this variability 
could come from random fluctuations in the stochastic-oscillator
system.  A first indication that random fluctuations are responsible
comes from that fact that most of the numbers in Table 1 agree within the
error estimates.

  To investigate this further, a long Monte Carlo time series was 
generated for a stochastic-oscillator process 
with the same parameters as the fit to
the observed NINO3.4 timeseries for 1951--1995, and
a sample of 500 independent 15-year long runs of this run
was analyzed.  In the analysis, the mean seasonal cycle of each
run was removed, except for the calculation of the bias.

\begin{figure}
\par
\vspace*{1mm}
\centerline{\hspace*{0mm}\hbox{
\psfig{figure=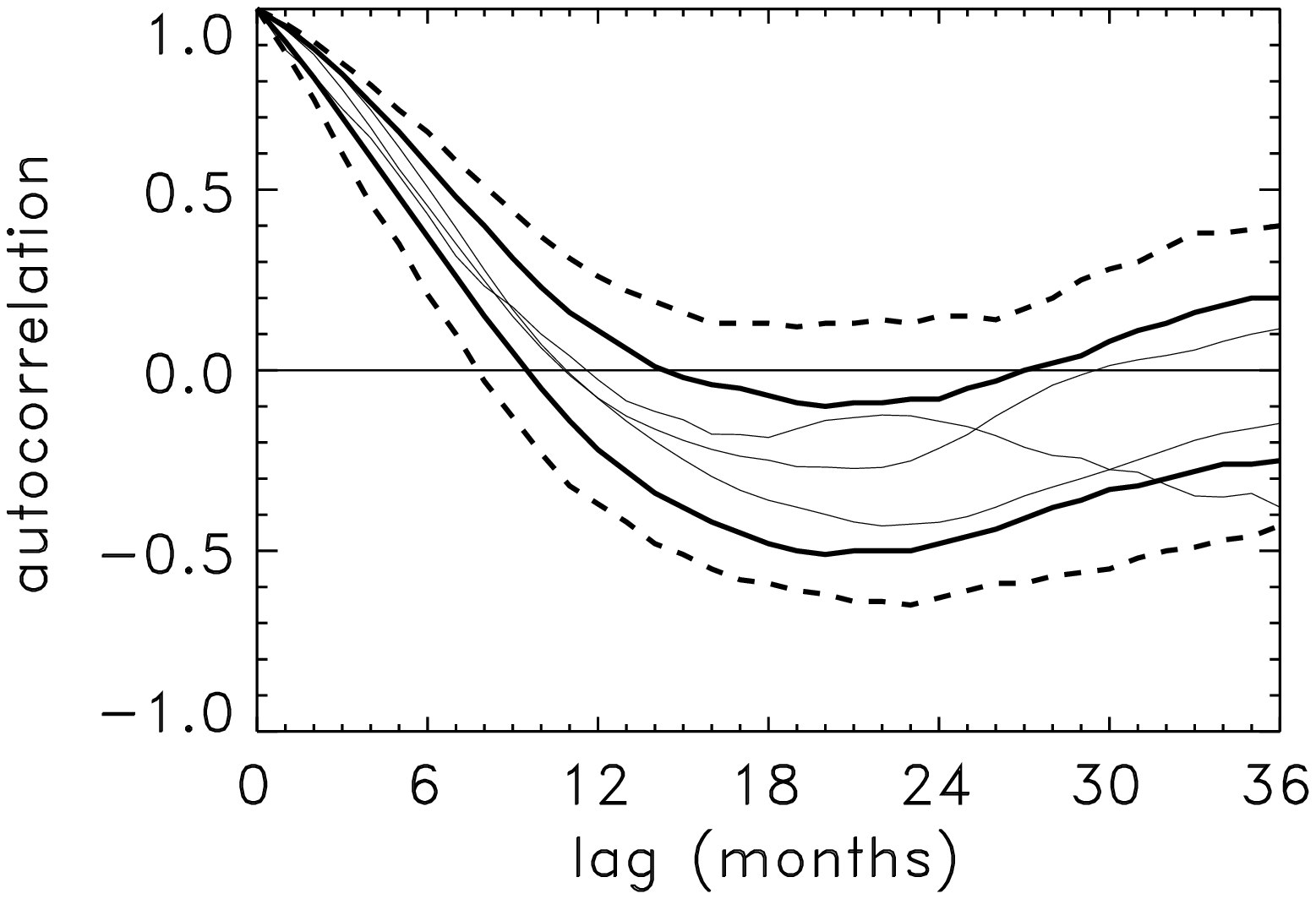,width=10.0cm,angle=0}
}}
\par
\addtolength{\baselineskip}{-2mm}
{\small
\setlength{\parindent=5mm}             
\indent {\sc Fig.\ 5.}
Spread of the autocorrelation in a sample of 15-year long timeseries
generated by a stochastic oscillator fitted to the observed
timeseries 1951-1995.
At a given lag, the autocorrelation falls in 68\%  (95\%)
of the cases between the thick lines (thin lines).
Also shown are the observed autocorrelations of the same three 15-year long
subperiods as in Fig.\ 1 (thin lines).
}

\end{figure}


In this sample, values for the bias of $0.00 \pm 0.13$ and 
for standard deviation of 
$0.68 \pm 0.12$ were found.  
So the difference in bias between
1951-1980 and 1981-1995 represents a two standard deviation
effect.
How much the autocorrelation functions of the sample varied
is shown in Fig.\ 5.
For comparison, the observed autocorrelations over the three
15 year periods of Fig.\ 1 are shown as well.
Generally, they fall within
the 68\% ranges of Fig.\ 5.  Close inspection reveals that only 
for lag 1 over the subperiod
1951-1966 the observed autocorrelation touches the 95\% range,
perhaps due to larger observational errors during this period,
cf. Fig.\ 4.

ARMA parameters were fitted to each member of the sample.
The variation in the parameters was large,
which is not surprising in view of the fact that a 15 
year period cannot contain the equivalent of many independent
estimates of a 4 year period.
The median value and the 68\% range were
$T = 47^{+17}_{-8}$ months, 
$D = 21^{+12}_{-11}$ months and
$k = 0.90^{+0.09}_{-0.13}$.
  
  For a longer period of 45 years, the variation in a 500
member sample is reduced to
$T = 47 ^{+8}_{-5} $ months, 
$D = 18 ^{+7}_{-5} $ months and
$k = 0.87 \pm 0.06$.

The forecast skill varies accordingly.  Using the ``correct''
parameters to forecast the index, 
the lead time at which the skill had dropped to $0.6$ 
was $\rm{FS}= 6 \pm 2$ months for the sample of 15-year long
runs and $\rm{FS}= 6 \pm 1$ months for the 45-year long
runs.

From the above Monte Carlo results follows 
that the observed decadal variability is similar to 
what one can expect if ENSO is a stochastic oscillator 
driven by short-term variability, although the observed change in bias
from 1951-1980 compared to 1981-195 is rather large.

\section{Seasonal effects \label{sec:SEA}}

In the stochastic oscillator discussed so far, El Ni\~{n}o episodes evolve
completely independent from the seasonal cycle. 
Observations indicate this is not the case.  

The variance of December anomalies is about three times 
as much as the variance of April anomalies.    
Perhaps part of the explanation is that
the mean thermocline in December is much more shallow
than in April, making SST much more sensitive to 
anomalies of the thermocline depth (Kleeman 1993)
in December than in April.
Another part of the explanation is that the noise is not    
stationary.  Here the seasonal motion of the ITCZ is likely
to play a role (Philander 1990, Tziperman et al.\ 1997).

Due to chance, there will also be
differences  between the variances of the calendar months 
in the stochastic-oscillator runs.
A series of 500 Monte Carlo runs of 45 years length was made
to examine the size of the differences.  
Only in 8 cases 
the ratio of the smallest variance to the largest was 
smaller than $0.5$, and it was never smaller than $0.35$.
The median value was $0.7$. 
This is to be compared to the observed value, which is as small as $0.31$.
This analysis confirms that a stationary stochastic oscillator cannot
give a seasonal dependence of the variances of the anomalies that
is as large as observed.

Another seasonal phenomenon in the observations is    that the
autocorrelation is not stationary.   The correlation between a January 
anomaly and the anomaly 6 months earlier is about $0.8$, while
the correlation between an August anomaly and the anomaly 6 months
before is only about $0.2$.  In the stochastic oscillator of section
{\ref{sec:MOD}} the lag-6 correlation is about $0.5$, 
independent of the season.

A qualitative explanation of the seasonal dependence of the correlation
is that when the anomaly variance is small,
the influence of the noise will be large, keeping all other things equal
(Philander 1990).
This means there is a ``spring barrier'' in predictability, caused
by the months around April, when the anomaly variance is low.

\begin{table}[b]
\begin{center}
\begin{tabular}{|l|llll|}
\hline  
    &  \hspace*{2mm} T   &  \hspace*{2mm} D   &  \hspace*{4mm} k     &  FS \\
\hline  
1951--65 & $ 90 ^{+\infty}_{-25}$ & $ 15 ^{+7}_{-3} $ & $ 1.00 ^{+0.00}_{-0.03} $ &   5   \\
1966--80 & $ 48 ^{+12}_{-7} $ & $ 18 ^{+11}_{-7} $ & $ 0.87 ^{+0.05}_{-0.08} $ &   7   \\
1981--95 & $42  ^{+ 7}_{-6} $ & $ 21 ^{+20}_{-10} $ & $ 0.79 ^{+0.10}_{-0.15} $ &   7.5 \\
1951--95 & $ 55 \scriptstyle { \pm 8 } $ & $ 19 \scriptstyle { \pm 5 } $ &
 $ 0.90 \scriptstyle { \pm 0.04 } $ &   6.5 \\
\hline  
\end{tabular}
\end{center}

\par
\addtolength{\baselineskip}{-2mm}
{\small
\setlength{\parindent=5mm}
\indent{\sc Table 2.}
Parameters of a stochastic-oscillator fit to 
the standarized NINO3.4 index.  
Period T (months), decay time scale D (months),
ARMA parameter k, 
and lead FS (months) at which the anomaly correlation skill of 
stochastic-oscillator forecasts for the standarized anomaly drops to $0.6$.
}

\end{table}

Again, it is examined how large the differences in correlation 
are that occur by chance in stochastic-oscillator runs.  In each
of the runs, the largest difference in correlation between pairs
with different starting months but the same lag was determined.
In the figures reported here, the maximum permitted lag was 6 months.
Of 500 45-year long runs of a stochastic 
oscillator 
the median of this difference was 0.25, 16\% had a difference
larger than 0.35,
and no one had a difference larger than 0.55.
For 15-year runs, the differences were much larger, in this case 
the median difference was as large as 0.45!
This analysis indicates that it is unlikely that the
stochastic oscillator produces
the observed seasonal dependence of the correlation, but also that it
is quite hard to make a reliable estimate of the seasonal dependence
of the correlation from observations.

The above results show that the stationary
stochastic oscillator model of section \ref{sec:MOD} is
incomplete because it lacks
seasonal dependence in variance    and correlation.

In principle, the extension is clear: make all parameters 
dependent on the season, and repeat the analysis to get the best fit to
the observed seasonal dependence.
This will not be attempted here. 
Instead, the following simple trick is used.
The analysis of section \ref{sec:MOD} is redone
on data scaled to unit variance month-by-month.  
Stochastic-oscillator parameters from
fits to these standarized anomalies are
given in Table 2.  The values are similar to those for fits to the full
NINO3.4 anomalies in Table 1.

A stochastic oscillator model of the standarized anomalies is
equivalent to a seasonal stochastic oscillator 
which is obtained by 
multiplying the  r.h.s. of    (\ref{eq:stoch}) by a 
factor that depends on the calendar month, the products of these 
factors being unity.  
This makes that for the NINO3.4 seasonal stochastic oscillator
the deterministic part grows in boreal fall and decays relatively fast in
early spring, giving rise to the observed seasonal dependence
of the variance.
Also, it implies that the
noise is larger if the deterministic part favours growth, thus
neglecting the seasaonal dependence of the correlation.

\begin{figure}
\par
\vspace*{-2mm}
\centerline{\hspace*{0mm}\hbox{
\psfig{figure=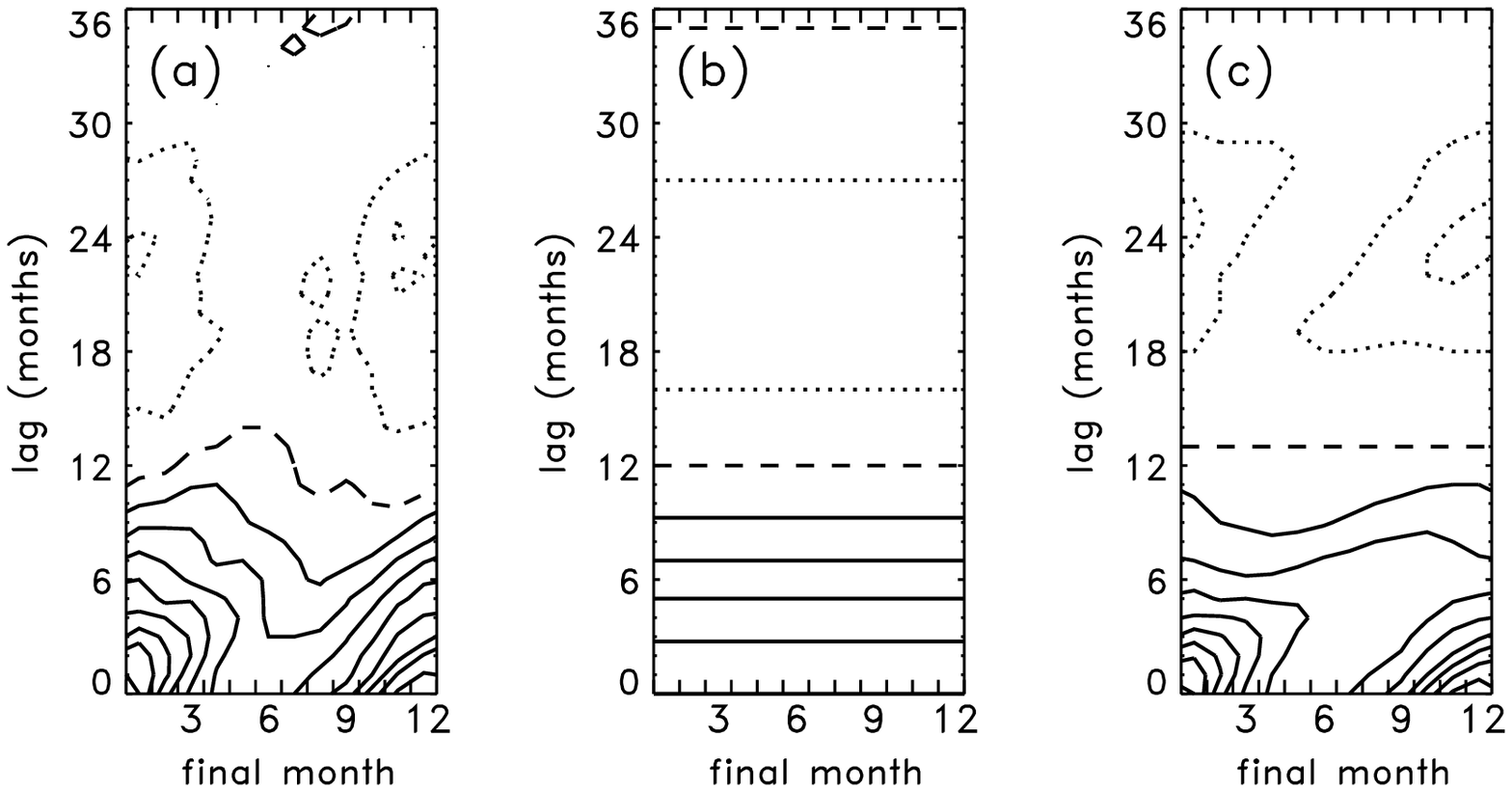,width=16.0cm,angle=0}
}}
\par
\addtolength{\baselineskip}{-2mm}
{\small
\setlength{\parindent=5mm}             
\indent {\sc Fig.\ 6.}
Seasonal autocovariance as function of lag and final month.
(a) Observed covariance of the NINO3.4 index over the period 1951-1995.
(b) Covariance of a stationary stochastic oscillator fitted to 
observations.               
(c) Covariance of a stochastic oscillator
with seasonal variances but stationary correlations, fitted to
observations.               
Contour interval is $0.1 {\mathrm K}^2$.
}
\end{figure}


Although the above modification  leaves the correlation unchanged and
independent of the season, it is a great improvement of the covariance.  
In Fig.\ 6. the observed covariance as a function of lag and final month
is compared to that of a stationary stochastic-oscillator fit and to 
that of a fit of a stochastic oscillator with seasonal variance.
Note that although their  correlations 
are almost identical and independent of the season, 
the second and the third panel look less similar than
the first and the third.  

Actually, the anomaly correlation of the seasonal
stochastic-oscillator forecast for the full (destandarized) 
anomalies is somewhat 
larger than the skills of Table 2.
This is because high-variance months are generally better
predictable than low-variance months in terms of anomaly correlation,
due the observed seasonal dependence of the correlations.
This applies to persistence skill as well: forecasts
based on persistent standarized anomalies have more skill
than forecasts based on persistent anomalies.

\section{Forecast skill of the stochastic oscillator \label{sec:FOR}}

The stochastic oscillator model can be used for making El Ni\~{n}o forecasts.
This is an interesting test.
According to the model, one mode dominates, and predictability is 
inherently limited by the noise.  
Although the representation of the
noise process in ({\ref{eq:stochx}}) may be a oversimplification, it
is unlikely that this could make a difference of more than one or two  
months in skill. 
Also using more data, which would give a better determination
of $x$ and $y$ in ({\ref{eq:stoch}}), will give only a limited increase 
in skill.
This means that if the stochastic-oscillator forecasts would have much less 
skill than forecasts of more comprehensive models, then
something should be wrong with the concept of the
El Ni\~{n}o stochastic oscillator.

\begin{figure}
\par
\vspace*{-15mm}
\centerline{\hspace*{0mm}\hbox{
\psfig{figure=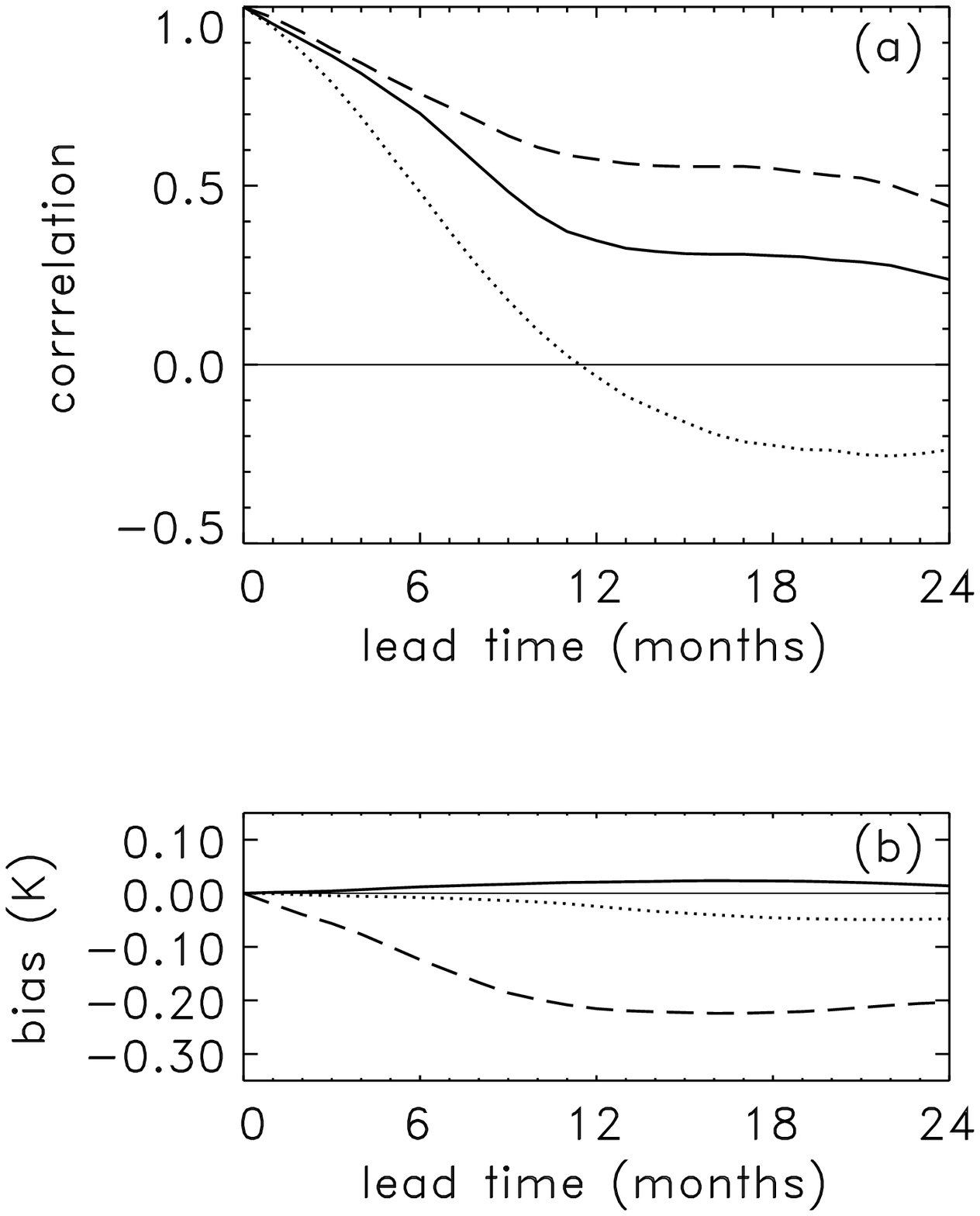,width=8.0cm,angle=0}
}}
\par
\addtolength{\baselineskip}{-2mm}
{\small
\setlength{\parindent=5mm}             
\indent {\sc Fig.\ 7.}
A posteriori forecast skill of the stochastic oscillator in terms of anomaly
correlation (a) and bias (b) over the periods 1956-1995 
(solid lines) and 1982-1993 (dashed lines).  The persistence
skill over the period 1956-1995 is indicated by the dotted lines.
A posterio means that for all forecasts the same model parameters were
used, determined from all data over the period 1956-1995.
}
\end{figure}


Given the time scales $ \gamma^{-1}$ and $ 2 \pi \omega^{-1}$, 
and the phase angle $\alpha$,
one can make a Kalman filter based on ({\ref{eq:stoch}}). 

The stochastic-oscillator variable $x$ is identified with the     
standarized NINO3.4 index. 
The observation error in the NINO3.4 index is neglected, because
the errors in the observation are small compared to the noise variance
found in section {\ref{sec:MOD}} and anyway part of the error may 
have been accounted for by the fit.  
During the assimilation, an estimate is obtained for the variable $y$ 
that is not observed, toghether with an estimate of the uncertainty
in $y$.  For the linear system of    ({\ref{eq:stoch}}), the uncertainty
in $y$ does not depend on the observed values of $x$ and converges 
to an asymptotic value in a few years.

The analysis step of the model variables and the Kalman gain is given by 
\begin{eqnarray}
     x^{\mathrm ana}_i & = & x^{\mathrm obs}_i  \\ \nonumber
     y^{\mathrm ana}_i & = & y^{\mathrm for}_i + 
                      {{(P_{12}^{\mathrm for})_i} \over {({P_{11}^{\mathrm for})_i}}} 
                                (x^{\mathrm obs}_i-x^{\mathrm for}_i ) \\ \nonumber
     (P_{11}^{\mathrm ana})_i & =  & 0  \\ \nonumber
     (P_{12}^{\mathrm ana})_i & =  & 0  \\ \nonumber
     (P_{22}^{\mathrm ana})_i & = & (P_{22}^{\mathrm for})_i - 
                 {     (P_{12}^{\mathrm for})_i^2 \over  
                       (P_{11}^{\mathrm for})_i }   \; .
\end{eqnarray}
The forecast step is  
\begin{eqnarray}
     x^{\mathrm for}_{i+1} & = & a   x^{\mathrm ana}_i - b y^{\mathrm ana}_i \\ \nonumber
     y^{\mathrm for}_{i+1} & = & b   x^{\mathrm ana}_i + b y^{\mathrm ana}_i \\ \nonumber
     (P_{11}^{\mathrm for})_{i+1} & = & 
                    a^2 (P_{11}^{\mathrm ana})_i - 2 a b (P_{12}^{\mathrm ana})_i +
                    b^2 (P_{22}^{\mathrm ana})_i + q_{11}\\ \nonumber 
     (P_{12}^{\mathrm for})_{i+1} & = & 
                    ab (P_{11}^{\mathrm ana})_i + (a^2-b^2)
           (P_{12}^{\mathrm ana})_i - ab (P_{22}^{\mathrm ana})_i + q_{12}\\ \nonumber
     (P_{22}^{\mathrm for})_{i+1} & = & 
                    b^2 (P_{11}^{\mathrm ana})_i + 2 a b (P_{12}^{\mathrm ana})_i + 
                    a^2 (P_{22}^{\mathrm ana})_i + q_{22}  \; ,
\end{eqnarray}
with the  $q_{ij}$ the noise covariances, see {\ref{sec:APP}}.
In case $(q_{12})^2 = q_{11} q_{22}$, 
when in ({\ref{eq:stoch}}) $  \eta_i /  \xi_i $ is a constant,
one has that
$\lim_{i \rightarrow \infty } (P_{22}^{\mathrm ana})_i = 0$
for the Kalman estimate of the error variance in $y$,
and the above equations simplify considerably.
Suitable starting values for $y$
and $P_{22}$ are
$y_0 = ( \langle  xy \rangle / \langle xx \rangle ) x_0$ and
$(P_{22})_0 = \langle y^2 \rangle - \langle xy \rangle ^2 / 
\langle x^2 \rangle$.

    Forecasts can be made starting from an analysis of $x_i$ and $y_i$ 
and $(P_{22})_i$.
The Kalman equations  give forecasted values for $x$ and $y$, 
and also forecasted values for the uncertainties in $x$ and $y$.
An alternative is making an ensemble of forecast runs, 
using ({\ref{eq:stoch}}) with different noise for each ensemble member.
Finally, the forecasted values for $x$ are destandarized to get
the full anomalies.

The Kalman filter was tried on the NINO3.4 index, because
the NINO3.4 index is more persistent and  easier to predict than 
the NINO3 index.

In a first experiment, forecasts 
over the period 1956--1995 were made with a model
fitted to the same period.  
Because the model parameters were determined 
using the timeseries
over the full period, these are not really forecasts in the strict
sense that no future information is used, and the skill is   
overestimated substantially.  In principle, the same applies to 
persistence forecasts based on a climatology calculated over
the full period too, but in that case the effect is only marginal.

The skill as measured by the anomaly correlation and the bias
of these ``a posteriori'' forecasts is shown in Fig.\ 7.  
The correlation
drops to about 0.6 after $7.5$ months.  This amounts to a
gain of $2.5$ month over simple persistence skill and of
$1.5$ month over ``standarized persistence'' skill (not shown).  
Also shown in Fig.\ 7 is that the skill was higher 
over the subperiod 1982--1993 than over   1956-1995.
However, for longer lead times, the bias over this subperiod is large.
This is because for longer lead times, the 
stochastic-oscillator forecasts relax to the climatology of the full period 
instead of to the subperiod mean.

\begin{figure}
\par
\vspace*{1mm}
\centerline{\hspace*{0mm}\hbox{
\psfig{figure=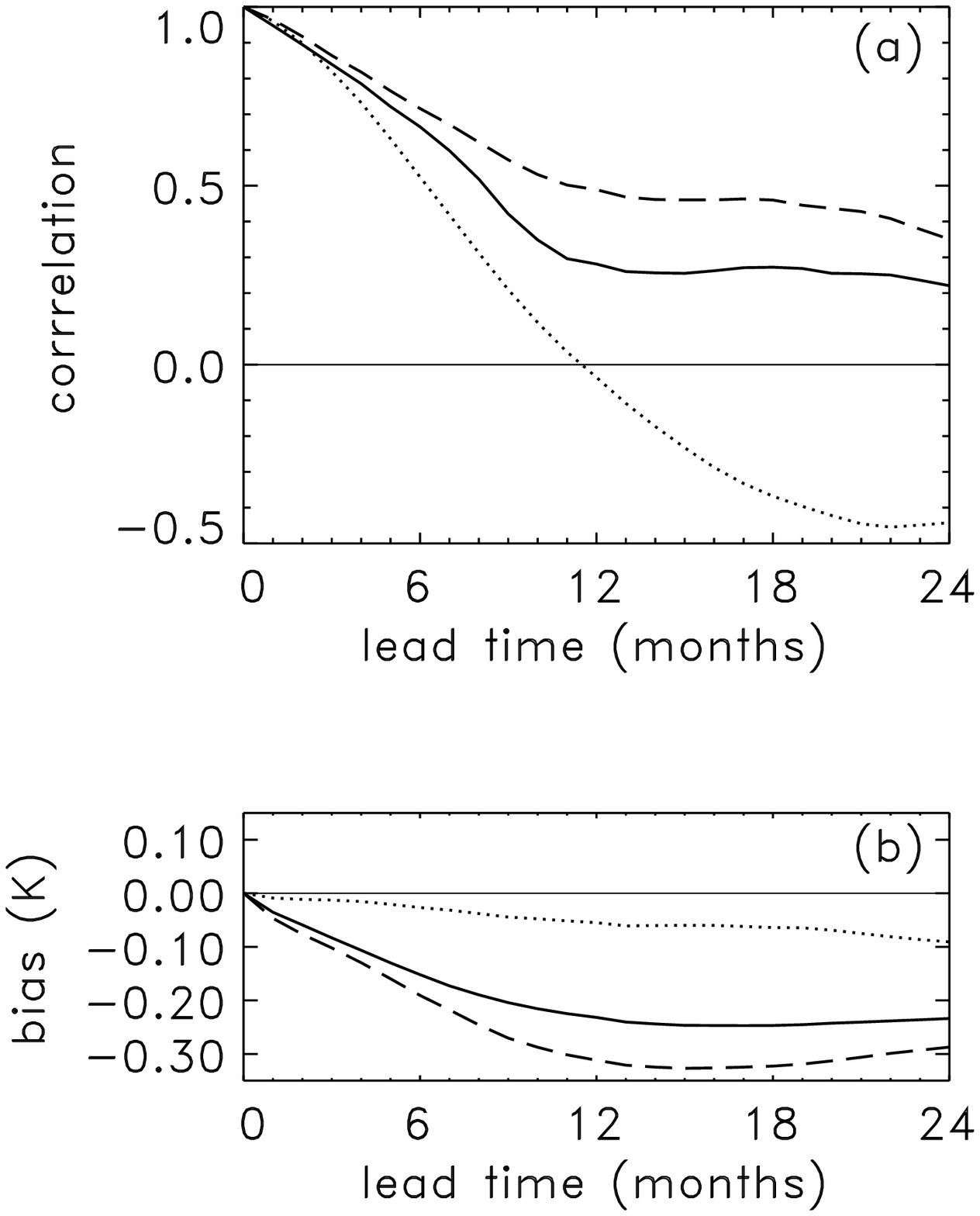,width=8.0cm,angle=0}
}}
\par
\addtolength{\baselineskip}{-2mm}
{\small
\setlength{\parindent=5mm}             
\indent {\sc Fig.\ 8.}
Quasi-operational forecast skill of the stochastic oscillator in terms of anomaly
correlation (a) and bias (b) over the periods 1976-1995 
(solid lines) and 1982-1993 (dashed lines).  The persistence
skill over the period 1982-1993 is indicated by the dotted lines.
Quasi-operational means that for each forecast, the model parameters were 
determined using only data from before the forecast period.
}
\end{figure}


In another experiment, quasi-operational forecasts
were made for the period 1976--1995.  
The model parameters 
(biases, variances, stochastic-oscillator parameters) 
for each forecast were determined using 
only data from $1956$ to the starting month of the forecast.
The skill of these quasi-operational forecasts is shown in Fig.\ ~8.
Also shown in Fig.\ 8 is the skill of the quasi-operational
forecasts over the subperiod 1982-1993.
As one sees, the quasi-operational skill is somewhat lower than
the a posterio skill.  Over 1982-1993, the anomaly 
correlation drops to 0.6 after 8.5 months, and the bias is
about $-0.3$K for longer lead times.

The reader may check the forecast of Fig.\ 9, which was made 
just before this paper was submitted.  It shows the Kalman Filter forecast
from February 1997, toghether with a 32-member ensemble forecast.

A comparison of two-season lead forecasts by Barnston et al.\ (1994) 
concentrated on the relatively well-predictable sub-period 1982--1993.  
The anomaly correlation 
of a quasi-operational two-season lead forecast as defined by
Barnston et al., which roughly corresponds to the anomaly correlation
at eight months as presented here, is $0.65$ for the quasi-operational
forecasts.  
This is comparable to the skill of the
models considered in Barnston et al.\ and significantly better than
the anomaly-correlation skill of persistence, 
which is $0.33$ for the period considered.

The above results indicate that the stochastic oscillator is 
quite capable of succesful predicting the main ENSO mode.

\begin{figure}
\par
\vspace*{1mm}
\centerline{\hspace*{0mm}\hbox{
\psfig{figure=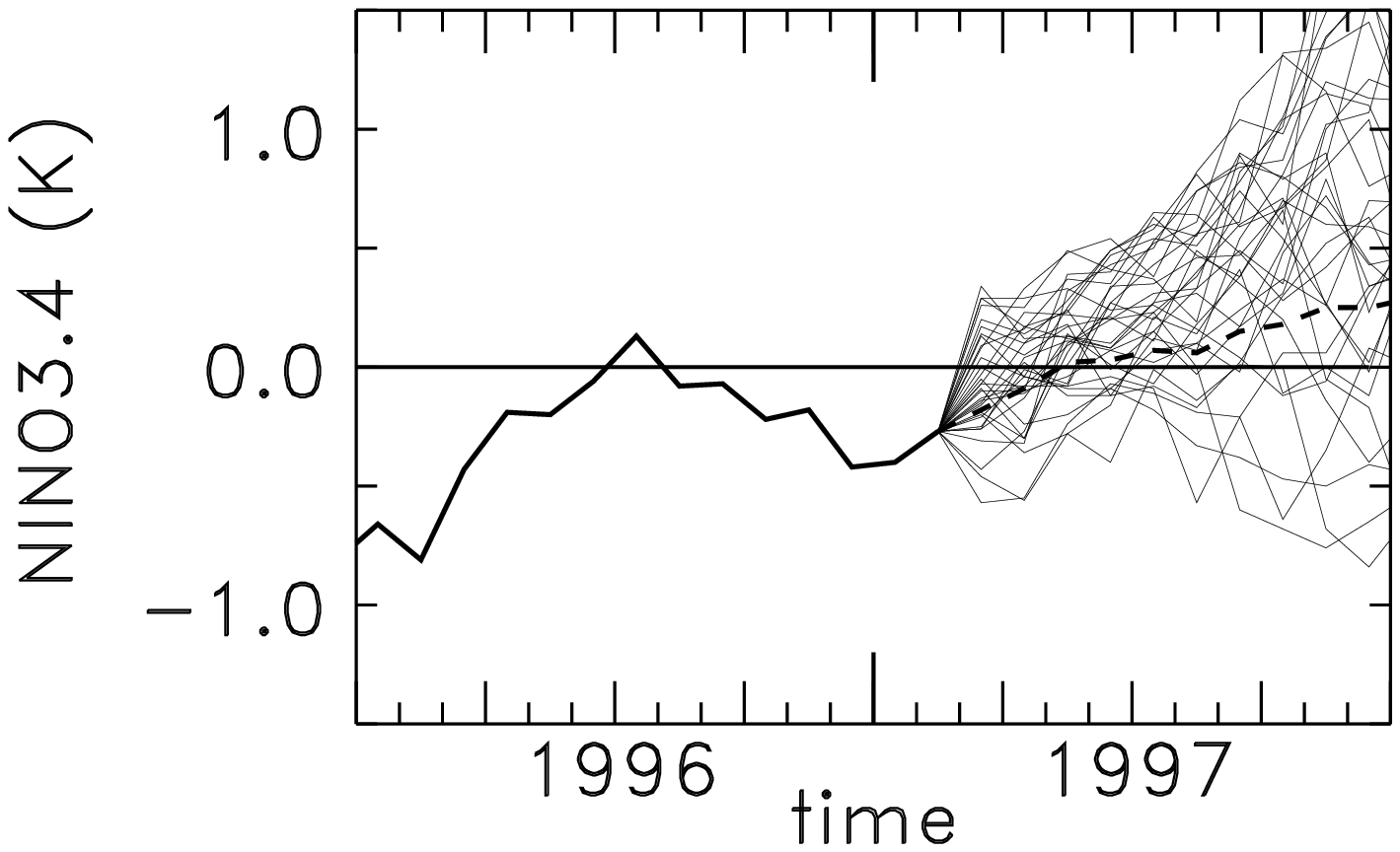,width=12.0cm,angle=0}
}}
\par
\addtolength{\baselineskip}{-2mm}
{\small
\setlength{\parindent=5mm}             
\indent {\sc Fig.\ 9.}
Stochastic-oscillator forecast from February 1997.  Observations until
February 1997 are denoted by the thick line, the 
stochastor-oscillator Kalman Filter
forecast by the thick dashed line.  
A 32-member ensemble of stochastic-oscillator forecasts is plotted
as thin lines.
}
\end{figure}


\section{Discussion and conclusions \label{sec:DIS}}

Two loosely connected themes have been treated in this paper.

The first theme is that of the dynamical role of ``external noise''.
Noise can not only be important for the irregularity of El Ni\~{n}o, 
but also could excite El Ni\~{n}o's.  
The picture of El Ni\~{n}o discussed here 
is that it could be a low-dimensional stable system excited by external noise.
This can be modelled by a stochastic-oscillator equation,
because El Ni\~{n}o is dominated by a single mode. 
The stochastic-oscillator parameters were shown to be compatible 
with the range of values
                usually assumed for the delayed-oscillator parameters.
Of course, a stable mean
process in the stochastic oscillator corresponds to a stable
linearized delayed oscillator.  But 
the parameters found for the stochastic oscillator indicate a
{\em locally} unstable process in the delayed oscillator, and
by rather small changes of parameters
one can switch from a stable to an unstable delayed oscillator.

The second theme is 
that El Ni\~{n}o's can be predicted rather well 
by an ARMA(2,1) (autoregressive-moving average) process because
the observed autocorrelation function of El Ni\~{n}o indicators can be
parameterized quite well by that of a stochastic oscillator.
The autocorrelation of a stochastic oscillator has {\em three} parameters: 
in addition to a period and a decay scale, there is a phase shift.
This provides a convenient reference in skill for more comprehensive models.
That the ARMA(2,1) model is that succesful is again because   
El Ni\~{n}o is dominated by a single mode.  

The connection is as follows.
If the stochastic oscillator is a good picture of El Ni\~{n}o, then
an ARMA(2,1) model (or a not too fanciful generalisation)
should be rather succesful, although the reverse is not true.

So the stochastic-oscillator picture has passed two important tests
--- compatibility with the delayed-oscillator picture, and 
a forecast model based on the stochastic oscillator performs well ---
but remains far from proven.

It is amusing to note that non-linear delayed-oscillator models without
external noise need
an unstable basis state while a linear stochastic oscillator needs
a stable one.  Even more so because observations indicate that
the basic state is part of the year stable and part of the year
unstable.   
It could be that what is considered noise in the fits
to the stochastic oscillator is actually the manifestation of
fluctuations intrinsic to a low-dimensional chaotic El Ni\~{n}o
systen.  To resolve this question falls outside the scope of this
article, because
the observed timeseries is far
too short to make a distinction between chaos and noise (Tziperman
et al.\ 1994).

   The seasonal influence on El Ni\~{n}o is so large that I
could not neglect it altoghether.  It can be represented in a crude
way by doing as if the stochastic oscillator applies to the
standarized anomalies.

   The form of the stochastic-oscillator equations leads naturally to
a Kalman Filter for forecasting anomalies.  The observed NINO3.4   
indices are the only input.
This filter 
makes rather good forecasts of the main mode of the ENSO system,
comparable to the skills quoted in the study of 
Barnston et al. (1994).
There is no reason to believe that one cannot do better and
gain some precious months in forecast skill.
Indeed, recent results, in particular of models that assimilate 
sub-surface information (Kleeman et al.\ 1995, Ji et al.\ 1996), 
seem to point to an 
improvement in model skill in recent years.

Even within the context of the stochastic oscillator, improvement in
forecast skill could come from more comprehensive models.
Seasonality might be represented better,
data-assimilation might give an better   estimate of the
two deterministic degrees of freedom of the system, and the noise 
might be better represented and partly predictable.
Also, going slightly beyond the stochastic-oscillator picture, it   
might very well be that the noise variance could be dependent on
El Ni\~{n}o.  

If models as the stochastic oscillator are tuned on relatively short
testing periods, forecast skill may be overestimated considerably. 

The stochastic oscillator exhibits substantial decadal variability.
This applies not only to the apperent mean state and that some
periods are more active than others.
It also applies to predictability: even with a ``perfect'' model,
during some periods forecasts are much more succesful than  
during others.

   {\em Acknowledgements}  I thank Geert Jan van Oldenborgh and
Gerbrand Komen for many discussions and Adri Buishand for providing me with
a maximum-likelihood algorithm for determining the parameters of an ARMA 
process.

\renewcommand\appendix{\par
  \setcounter{section}{0}%
  \setcounter{subsection}{0}%
  \renewcommand\thesection{Appendix \Alph{section}}}

\setcounter{equation}{0}
\renewcommand\theequation{A\arabic{equation}}

\appendix

\section{Properties of the stochastic oscillator}\label{sec:APP}

The simplest form of the stochastic oscillator is the
complex Langevin equation
\begin{equation}
\label{eq:lang}
                    \dot{z} = - c z + \zeta(t) \,
\end{equation}
for a complex variable $z$, with a complex constant $c$ 
and a noise term $\zeta$.  
The noise $\zeta$ is not specified as a given
function of time, but it assumed that its average 
$\langle  \zeta \rangle   = 0 $, 
that $\zeta(t)$ is distributed according a Gaussian law 
and that its second moment $ \langle \zeta(t) \zeta(t-\tau ) \rangle  $ 
is given.  
The autocorrelation of $\zeta$ should go to zero at times much smaller 
than the mean time scale $\mid c \mid ^{-1}$, 
for a meaningful separation of mean and stochastic processes to be 
possible.     ({\ref{eq:lang}}) has the formal solution
\begin{equation}
\label{eq:langa}
                    z = z_0 e^{-ct} + e^{-ct} \int_0^t
                            d  \tau \, e^{c \tau} \zeta ( \tau ) \; .
\end{equation}
A clear account of the theory of the real Langevin equation can be 
found in Balescu (1975).  

In this paper, the discretized form of    ({\ref{eq:lang}}) is 
considered:
\begin{eqnarray}
\label{eq:stoch1}
     x_{i+1} & = & a x_i  - b y_i + \xi_i    \nonumber \\
     y_{i+1} & = & b x_i  + a y_i + \eta_i    \,
\end{eqnarray}
for two real variables $x$ and $y$, real constants $a$ and $b$ 
and noise terms
$\xi _i$ and $\eta _i$.  From    ({\ref{eq:stoch1}}) follows 
the following equation in terms of $x$ alone:
\begin{equation}
\label{eq:stochy}
   x_{i+1} = 
     2 a x_i - (a^2+b^2) x_{i-1} + \xi_i - a\xi_{i-1} - b\eta_{i-1} \; .
\end{equation}
In the following, the timescales $\gamma^{-1}$ 
and $2 \pi  \omega^{-1}$ will be used.  They are defined by 
\begin{equation}
      a + i b = e ^{- \gamma\Delta t} \, e ^{i \omega \Delta t } \; ,
\end{equation}
with $\Delta t$ the timestep.

Next, the assumption is made that the noise 
between different timesteps in    ({\ref{eq:stoch1}})
is not correlated.  Going back to a complex variable and using formal 
solutions analoguous to    ({\ref{eq:langa}}),  the following 
expressions for the variances of the variables $x$ and $y$ in term of 
the variances of the noise can be obtained:
\begin{eqnarray}
\label{eq:varia}
\langle  x^2 \rangle   & = & 
{ 1 \over 2}  {{ (q_{11} + q_{22} ) } \over { 1 - e^{ -2 \gamma \Delta t}}}  
  + { 1 \over 2}  {{ ( 1 - 2 e^{-2\gamma\Delta t} \cos(2 \omega\Delta t) ) \, 
                                                (q_ {11}-q_{22})} \over 
     { 1 - 2 e^ { -2 \gamma\Delta t } \cos(2 \omega\Delta t) \,+ 
                                           \, e^{-4\gamma\Delta t}}} 
                                                        \\ \nonumber
  &  &  
  - {{ e^{-2\gamma\Delta t} \sin(2 \omega\Delta t)\, q_ {12} } \over 
     { 1 - 2 e^ { -2 \gamma\Delta t } \cos(2 \omega\Delta t) \,+ 
                                           \, e^{-4\gamma\Delta t}}} 
                                                       \\ \nonumber
\langle   y^2 \rangle   & = & 
   { { q_{11} + q_{22} } \over { 1 - e^{ -2 \gamma\Delta t } }} \; 
   - \; \langle   x^2 \rangle  
                                                        \\ \nonumber
\langle   xy \rangle    & = & 
{ 1 \over 2}  {{ e^{-2\gamma\Delta t} \sin(2 \omega\Delta t)\, 
                                              (q_ {11} - q_{22})} \over 
     { 1 - 2 e^ { -2 \gamma\Delta t } \cos(2 \omega\Delta t)\,+ 
                                           \, e^{-4\gamma\Delta t}}} 
  + {{( 1 - 2 e^{-2\gamma\Delta t} \cos(2 \omega\Delta t )) \, 
                                                       q_ {12} }  \over 
     { 1 - 2 e^ { -2 \gamma\Delta t } \cos(2 \omega\Delta t)\,+ 
                                           \, e^{-4\gamma\Delta t}}} \; ,
\end{eqnarray}
with
\begin{eqnarray}
        q_{11} & = &  \langle  \xi_i \xi_i \rangle  \\ \nonumber
        q_{22} & = &  \langle  \eta_i \eta_i \rangle  \\ \nonumber
        q_{12} & = &  \langle  \xi_i \eta_i \rangle     \; .
\end{eqnarray}
The autocorrelation function of $x$, 
$\rho_k = \langle  x_{i+k} x_i \rangle  / \langle  x_i x_i \rangle $ is given by
\begin{equation}
\label{eq:autocoa}
 \rho_k = 
                           e^{- k \gamma \Delta t} 
               \cos ( k \omega \Delta t + \alpha ) / \cos \alpha \, ,
\end{equation}
\begin{equation}
 \tan \alpha = \langle xy \rangle  / \langle x^2 \rangle  \, .
\end{equation}
So in addition to the mean-process timescales $\gamma^{-1}$ and $\omega^{-1}$ 
there is a third parameter in the autocorrelation function, 
the phase shift $\alpha$.  The phase shift depends both on the mean and 
the noise parameters.  
In the special case that $q_{12}=0$ and $q_{11}=q_{22}$, one has
$\langle xy \rangle  =0$ and $\alpha=0$.  
In the limit $\omega \rightarrow 0$, $\omega \tan \alpha \rightarrow -r$,
one has 
$ \rho_k \rightarrow
      \exp (- \gamma k\Delta t ) ( 1 + r k \Delta t ) $.

    If one would make a POP (Hasselmann 1988) analysis of $x$ and $y$, 
or of observed fields dominated by a mode linearly related to $x$ and $y$, 
then the leading POP mode will have the stochastic-oscillator period 
and decay time scale.

Looking only at one variable, the stochastic oscillator reduces to 
what in term of autoregressive modelling is called an ARMA(2,1) 
(autoregressive-moving average) process:
\begin{equation}
\label{eq:stochxxx}
  x_{i+1} = 
      2 a x_i - (a^2+b^2) x_{i-1} + \epsilon_i - k \epsilon_{i-1} \; , 
\end{equation}
where the correspondence between the ARMA noise variance 
\begin{equation}
       q_{00}   =    \langle   \epsilon_i \epsilon_i \rangle   
\end{equation}
 and the noise parameters $q_{11}$, $q_{22}$ and $q_{12}$  is given by   
\begin{eqnarray}
(1+k^2) q_{00} & = &  (1+a^2) q_{11} + b^2 q_{22} + 2 a b q_{12} 
                                               \\ \nonumber
  k    q_{00} & = &  a q_{11} + b q_{12} \; .
\end{eqnarray}
Note that $k$ and $1/k$ are equivalent.
To one value of $k$ corresponds a curve of combinations of 
$q_{22}/q_{11}$
and $q_{12}/q_{11}$ which have the same variance and autocorrelation for
$x$ but differ in the variance and autocorrelation of $y$.

The power spectrum which corresponds to the autocorrelation function of
   ({\ref{eq:autocoa}}) is in the limit $\Delta t \rightarrow 0$ that of
a broad peak with a red tail:
\begin{equation}
\label{eq:power}
  S( \omega ' ) =     { 1 \over { 2 \pi } } 
     \left\{
 { { \gamma - \tan \alpha (\omega ' - \omega  ) } \over 
   {  ( \omega ' - \omega  ) ^2 + \gamma^2 } } 
   +  { { \gamma + \tan \alpha (\omega ' + \omega  )  } \over 
   {  ( \omega ' + \omega  ) ^2 + \gamma^2 } } 
     \right\}  \; .
\end{equation}

\pagebreak

\begin{center}      {\Large \sc references  }  \end{center}

\begin{list}{}{\setlength{\parsep}{0cm}\setlength{\itemsep}{0cm}
               \setlength{\itemindent}{-1.5cm}\setlength{\leftmargin}{1.5cm}}

{\small
\item          Balescu, R., 1975:       {\em Equilibrium and nonequilibrium
               statistical mechanics}, J. Wiley \& Sons, 742pp. 
\item          Balmaseda, M.A., D.L.T. Anderson, and M.K. Davey, 1994:
               ENSO prediction using a dynamical ocean model coupled 
               to statistical atmospheres.
               {\em Tellus,} {\bf 46A}, 497--511. 
\item          Barnett, T.P., M. Latif, M. Fl\"{u}gel, S. Pazan,
               and W. White, 1993:
               ENSO and ENSO-related predictability:  Part I - Prediction
               of equatorial Pacific sea surface temperature with a
               hybrid coupled ocean-atmosphere model.
               {\em J.\ Climate,} {\bf 6}, 1545--1566.
\item          Barnston, G.B., H.M. van den Dool, S.E. Zebiak, T.P. Barnett,
               Ming Ji, D.R. Rodenhuis, M.A. Cane, A. Leetmaa, N.E. Graham,
               C.F. Ropelewski, V.E. Kousky, E.A. O'Lenic, and R.E. Livezey,
               1994:
               Long-lead seasonal forecasts -- where do we stand?
               {\em Bull.\ Am.\ Met.\ Soc.,} {\bf 75}, 2097--2114.
\item          Battisti, D.S., and A.C. Hirst, 1989:
               Interannual variability in a tropical atmosphere-ocean model:  
               Influence of the basic state, ocean geometry and nonlinearity.
               {\em J.\ Atmos.\ Sci.,} {\bf 46}, 1687--1712.
\item          Cane, M.A., 1992:
               Tropical Pacific ENSO models:
               ENSO as a mode of the coupled system.
               {\em In:} Climate System Modeling, K.E. Trenberth ed.,
               Cambridge University Press, p583--614.
\item          Chang, P., L. Ji, and H. Li, 1997:
               A decadal climate variation in the tropical Atlantic Ocean
               from thermodynamic air-sea interactions.
               {\em Nature,} {\bf 385}, 516--518.  
\item          Eckert, C. and M. Latif, 1997:
               Predictability of a stochastically forced hybrid coupled
               model of El Ni\~{n}o.
               {\em J.\ Clim.,} {(\em accepted)}.
\item          Fl\"{u}gel, M. and P. Chang, 1996:
               Impact of dynamical and stochastic processes on the        
               predictability of ENSO.
               {\em Geophys.\ Res.\ Let.,} {\bf 23}, 2089-2092.
\item          Griffies, S.M., and K. Bryan, 1997:
               Predictability of North Atlantic multidecadal climate        
               variability.
               {\em Science,} {\bf 275}, 181--184.
\item          \rule[1mm]{10mm}{0.2mm}\hspace*{1mm}, and E. Tziperman, 1995:
               A linear thermohaline oscillator driven by
               stochastic atmospheric forcing.
               {\em J.\ Clim.,} {\bf 8}, 2440--2453.
\item          Gu D., and S.G.H. Philander, 1997:
               Interdecadal climate fluctuations that depend on exchanges   
               between the tropics and extratropics.       
               {\em Science,} {\bf 275}, 805--807. 
\item          Hasselmann, K., 1976:
               Stochastic climate models.  Part 1.  Theory.
               {\em Tellus,} {\bf 18}, 473--485.
\item          \rule[1mm]{10mm}{0.2mm}\hspace*{1mm}, 1988:
               PIP's and POP's: The reduction of complex dynamical systems
               using principal interaction and oscillation patterns.
               {\em J.\ Geophys.\ Res.,} {\bf 93}, 11015--11021.
\item          Hirst, A.C., 1986:
               Unstable and damped equatorial modes in simple coupled
               ocean-atmosphere models.
               {\em J.\ Atmos.\ Sci.,} {\bf 43}, 606--630.  
\item          Ji, M., A. Leetmaa, and V.E. Kousky, 1996:
               Coupled model predictions of ENSO during the 1980s and the   
               1990s at the National Centers for Environmental Prediction.
               {\em J.\ Clim.,} {\bf 9}, 3105--3120.
\item          Jin, F.-F., D. Neelin, and M. Ghil, 1994:
               ENSO on the devil's staircase.
               {\em Science,} {\bf 263},70--72.
\item          \rule[1mm]{10mm}{0.2mm}\hspace*{1mm}, 1997:
               An equatorial ocean recharge paradigm for ENSO.
               Part I: Conceptual model.
               {\em J.\ Atmos.\ Sci.,} {\bf 54}, 811--829.  
\item          Kleeman, R., 1993:
               On the dependence of the hindcast skill on ocean      
               thermodynamics in a coupled ocean-atmosphere model.
               {\em J.\ Clim.,} {\bf 6}, 2012--2033.
\item          \rule[1mm]{10mm}{0.2mm}\hspace*{1mm}, Moore and N.R. Smith, 1995:
               Assimiliation of sub-surface thermal data into an     
               intermediate tropical coupled ocean-atmosphere model.
               {\em Mon.\ Wea.\ Rev.,} {\bf 123}, 3103--3113.
\item          \rule[1mm]{10mm}{0.2mm}\hspace*{1mm}, and
               \rule[1mm]{10mm}{0.2mm}\hspace*{1mm}, 1997:
               A theory for the limitation of ENSO predictability due to
               stochastic atmospheric transients.
               {\em J.\ Atmos.\ Sci.,} {\bf 54}, 753--767.  
\item          Latif, M., 1987:
               Tropical ocean circulation experiments.
               {\em J.\ Phys.\ Oceanogr.,} {\bf 17}, 246--263.   
\item          Lau, K.M., 1985:
               Elements of a stochastic-dynamical theory of the long-term    
               variability of the El Ni\~{n}o/Southern Oscillation.
               {\em J.\ Atmos.\ Sci.,} {\bf 42}, 1552--1558.  
\item          Mantua, N.J., and D.S. Battisti 1995:
               Aperiodic variability in the Zebiak-Cane coupled    
               ocean-atmosphere model: air-sea interactions in the
               western equatorial Pacific.
               {\em J.\ Clim.,} {\bf 8}, 2897--2927.
\item          M\"{u}nnich, M., M. Latif, S. Venzke, and E. Maier-Reimer, 1997:
               Decadal oscillations in a simple coupled model.
               submitted to {\em J. Phys. Oceanogr.}
\item          Penland, C., and P.D. Sardeshmukh, 1995:
               The optimal growth of tropical sea surface temperature anomalies.
               {\em J.\ Clim.,} {\bf 8}, 1999--2024.
\item          Philander, S.G.H., 1990:  {\em El Ni\~{n}o, La Ni\~{n}a and 
               the Southern Oscillation.}
               Academic Press, 293pp.
\item          Reynolds, R.W., and T.M. Smith, 1994:
               Improved global sea surface temperature analyses using 
               optimum interpolation.
               {\em J.\ Clim.,} {\bf 7}, 929--948. 
\item          Schneider E.K., B. Huang, and J. Shukla,  1995:
               Ocean wave dynamics and El Ni\~{n}o.
               {\em J.\ Clim.,} {\bf 8}, 2415--2439.
\item          Tziperman, E., L. Stone, M.A. Cane, and H. Jarosh, 1994:
               El Ni\~{n}o chaos: Overlapping of resonances between the
               seasonal cycle and the Pacific ocean-atmospere oscillator.
               {\em Science,} {\bf 263},72--74.
\item          \rule[1mm]{10mm}{0.2mm}\hspace*{1mm}, Zebiak and M.A. Cane, 1997:
               Mechanisms of seasonal-Enso interaction.
               {\em J.\ Atmos.\ Sci.,} {\bf 54}, 61--71.   
\item          Zebiak, S.E.,  and M.A. Cane, 1987:
               A model El Ni\~{n}o-Southern Oscillation.
               {\em Mon.\ Wea.\ Rev.,} {\bf 115}, 2262--2278.   
}
\end{list}

\end{document}